\documentstyle[prb,twocolumn,psfig,aps,epsf]{revtex}
\begin{document}
\draft
\twocolumn[\hsize\textwidth\columnwidth\hsize\csname
@twocolumnfalse\endcsname
%
%
%

\title{Spectral density for a hole in an antiferromagnetic stripe phase}

\author{P. Wr\'obel$^1$ and R. Eder$^2$}

\address{$^1$ Institute for Low Temperature and Structure
Research,
P. 0. Box 1410, 50-950 Wroc{\l}aw 2, Poland}
\address{$^2$ Institut f\"ur Theoretische Physik, Universit\"at =
W\"urzburg,
Am Hubland, 97074 W\"urzburg, Germany}

\date{\today}
\maketitle

\begin{abstract}
Using variational trial wave function based on the string picture we
study the motion of a single mobile hole in the stripe phase of
the doped antiferromagnet. The holes within the stripes are taken to be =
static,
the undoped antiferromagnetic domains in between the hole stripes
are assumed to have alternating staggered magnetization, as is suggested =
by
neutron scattering experiments. The system is described by the
$t-t'-t''-J$ model with realistic parameters and we compute the single
particle spectral density.
\end{abstract}


\vskip2pc]
\narrowtext

\section{Introduction}
 The experimental discovery of static
stripe-like structures in some cuprate materials and
nickelates\cite{Tranquadaetal} has recently attracted considerable
attention. This is particularly remarkable for the cuprate
superconductors because stripe-like hole arrangements have been
predicted long ago by Hartree-Fock calculations for the $d$-$p$
model\cite{ZaanenGunnarson} as well as for the single-band Hubbard
model\cite{PoilblancRice} - these would then in some sense be the
first successful theoretical predictions ever made for these
materials. So far, there appears to be experimental evidence for
stripes in a CuO$_2$ based material only for the concentration of
$x=3D1/8$, where the material does not superconduct. This might
suggest that the stripes in this case are a kind of
CDW-instability, stabilized predominantly by Coulomb interaction
between holes, which occurs only at this special concentration. On
the other hand, the rather precise scaling of the vector of
magnetic incommensurability in La$_{2-x}$ Sr$_x$CuO$_4$  with $x$
for hole concentrations $x \leq 15\%$ can be quite naturally
explained by assuming that fluctuating stripes with an
x-independent hole density of $1/2$ within the stripes and a mean
spacing of $\approx 1/2x$ between stripes do exist, which are
separated by N\'eel ordered domains with alternating staggered
magnetization.\\ In that sense, stripes could be a generic feature
of a doped Heisenberg antiferromagnet. Exact diagonalization (ED)
studies of the t-J model have shown evidence for stripe-like hole
correlations in the ground state\cite{prelovsekzotos}, as well as
low-energy excited states where the holes form stripe-like
arrangements\cite{tk}. These results cannot actually be compared
to the experimental situation, because the relatively small size
of clusters available for the ED technique does not allow to
accommodate a sufficiently large supercell as is observed
experimentally. This is possible, however, in density matrix
renormalization group calculations and this technique also has
produced strong evidence\cite{whitescalapino} for stripe-like hole
arrangements in the ground state, or at least in very low excited
states of the t-J model.\\ In the present paper we do not want to
address the question whether stripes really do form in the t-J
model or what is the energy responsible for their stabilization.
Instead, we take the existence of a stripe-like hole arrangement
for granted, and study the dispersion and spectral function of an
additional hole, which is moving through the stripe `background'.
In other words: we want to discuss the photoemission spectrum of a
striped phase. This seems an interesting question, because
recently angle resolved photoemission (ARPES) data for La$_{2-x}$
Sr$_x$CuO$_4$ have become available\cite{Inoetal}. This compound
is known to show stripe order for $x=3D1/8$ and the ARPES spectra
show a rather nontrivial evolution from the hole dispersion in the
insulator $(x=3D0)$ which is very similar to
Sr$_2$CuO$_2$Cl$_2$\cite{Wellsetal}, via the formation of `new
states' near $(\pi,0)$ or $(0,\pi)$ to a very different spectrum
for the superconducting phase $(x\approx 0.15)$. Our intention is
to check whether this can be reproduced by studying, in much the
same way as the hole motion in an antiferromagnet is being
studied, the motion of a single mobile hole in a static background
of spins and immobile holes, shown in Figure \ref{fig1}, which
form a stripe pattern rather than the conventional undoped N\'eel
state.\\ We used the t-J  Hamiltonian:
\begin{equation}
H =3D - \sum_{i,j} \left( t_{ij} \hat{c}_{i,\sigma}^\dagger
\hat{c}_{j,\sigma} + H.c. \right) + J \sum_{\langle i, j
 \rangle}  \left( \vec{S}_i \cdot \vec{S}_j
 -\frac{\hat{n}_i \hat{n}_j}{4}\right).
\end{equation}
Here $\hat{c}_{i,\sigma}^\dagger =3D c_{i,\sigma}^\dagger
(1-n_{i,\bar{\sigma}})$ are the standard conditional electron
creation operators, which avoid double occupancies, $\vec{S}_i$
the operators of electron spin, and $\hat{n}_i$ occupation
numbers. The hopping integral $t_{ij}$ between nearest neighbors
was taken to be $t=3D1$, that for 2$^{nd}$ nearest neighbors was
taken as $t'=3D-0.3$, the one between 3$^{rd}$ nearest neighbors was
$t''=3D0.3$. All other hopping integrals, as well as all exchange
constants for other than nearest neighbors were taken to be zero.
The nearest neighbor exchange constant was $J=3D0.4$.\\ The hole
motion in an antiferromagnet can be understood at least
qualitatively within the string
picture\cite{BulaevskiiNagaevKhomskii,Trugman}. A hole created in
a N\'eel state cannot propagate, because it leaves behind a trace
of misaligned spins. This means that the magnetic frustration in
the system increases to good approximation linearly with the
distance travelled by the hole, whence the hole experiences an
`effective potential' which traps it around the site where it has
been created. The wave function of the respective self-trapped
state can be obtained by diagonalizing the Hamiltonian in the
subspace of `string states' where the hole is connected to its
starting point $i$ by strings of defects with various length and
topology. True delocalization then is possibly only by virtue of
the transverse part of the Heisenberg exchange which can `heal'
the defects created by the hopping hole. In fact, if the first two
defects created by the hole are repaired by the transverse
exchange, this amounts to shifting the starting point of the
string to a 2$^{nd}$ or 3$^{rd}$ nearest neighbor of $i$. The
resulting next-nearest neighbor hopping band of width $\propto J$
is indeed well established by a variety of techniques, both
numerically and analytically\cite{dagoreview}.
One may expect quite similar processes to occur also in a
stripe-like background
such as the one shown in Figure \ref{fig1}. The holes in the
stripes thereby are to be considered as static.
Also in this case an added mobile hole cannot
propagate freely, because it creates frustration in the remaining
antiferromagnetic volume of the system. Again the spin-flip
parts will be necessary to enable propagation across the =
antiferromagnetic
stripes. The main difference then is that there may be additional
processes when the hole passes through the hole stripes,
and the fact that these hole stripes form something like
`$\lambda/2$-plates' for the propagating hole, because they
separate antiferromagnetic domains with opposite staggered
magnetization. As already mentioned the latter
property is an experimental constraint,
which is necessary to explain the incommensurate low energy peaks
in neutron scattering. The calculation for the stripe
background thus will be
somewhat more complicated than for the
case of the pure N\'eel state, but in principle the formalism
is precisely the same. In the following we outline the respective
calculation and present the results.
As will be seen below, the computed single-particle spectra show
some resemblance with the experimental results of Ino {\em et al.}.
\section{Spin polarons in the stripe structure}
Our goal is to discuss the motion of holes in the stripe structure
shown in Figure \ref{fig1}. In the following, we denote this state
as $|\Phi_0\rangle$. It corresponds to a concentration of static
holes of $1/8$$=3D$$12.5 \%$ and has been suggested by  experiments
\cite{Tranquadaetal}. The 32 sites contained in the rectangle in
Fig.\ref{fig1} form the unit cell of this structure. Some
theoretical studies suggest different structures
\cite{whitescalapino} but also indicate a tendency towards
formation of stripes. Charge or spin order within the stripes has
not actually been observed. The pattern depicted in Fig.\ref{fig1}
has been chosen for simplicity. The charge and spin fluctuations
within stripes might, in principle, be taken into account in
subsequent steps of approximation, but here we  restrict ourselves
to the simplest case possible, namely static holes and spins.
We stress again that our aim is not to discuss the stability of
the stripe structure, but rather to take it for granted, and
concentrate on the properties of an additional hole injected into
the stripe structure. We assume that its dynamics is represented
by the standard $t-J$ model, augmented by additional next-nearest
neighbor ($t'$) and next-next-nearest ($t''$) hopping terms. Such
terms are
necessary\cite{Nazarenko,wellscomp,EdOh3,belinicher,aligia2} to
explain the dispersion in the insulating CuO$_2$ plane as measured
by Wells {\em et al.}\cite{Wellsetal}.

As already mentioned, a mobile hole which has been created in an
ideal antiferromagnetic background behaves as if it is trapped in
a kind of potential well \cite{BulaevskiiNagaevKhomskii}. The
reason is that its movement creates strings of misaligned spins in
the antiferromagnetic structure and thus increases the magnetic
energy. These strings of defects, tend to restrict the motion of
the hole to the neighborhood of the cite where it has been
created. The notion of this effective potential well for the hole
is equivalent to the idea of `spin bags' which was introduced in
the context of the Hubbard model\cite{SchriefferWenZhang}.
Coherent motion of holes is possible due to relaxation of strings
which is enabled predominantly by the transverse part of the
Heisenberg exchange contained in the $t-J$ model. The motion of a
hole, then, has a twofold character. Because of $t >
J,t,t^{\prime},t^{\prime \prime}$, the hole will oscillate rapidly
(with a hopping rate $\propto t^{-1}$) in the neighborhood of the
site to which it is attached by the string and  only rarely (at a
rate $\propto J^{-1}$) the exchange part will shorten the string
of defects by two. This means that the starting point of the
string, and hence the center of gravity of the oscillatory motion,
is shifted by two lattice spacings.

In order to describe this twofold dynamics of holes we first construct
wave functions which describe a self-trapped
state of the hole.
We first introduce the operators
\begin{equation}
T_{\langle i,j \rangle}=3D\left( c^{\dag}_{i,\downarrow}
c_{j,\downarrow} + c^{\dag}_{j,\uparrow} c_{i,\uparrow} \right).
\end{equation}
In an arbitrary spin background they shift holes by one lattice
spacing. By consecutively applying the $T_{\langle i,j \rangle}$
to a the state $\hat{c}_l |\Phi_0\rangle$, we generate a family of
string states. The strings of defects connect the hole to the site
$l$. We denote such a string state by $|l,{\cal P}_l \rangle$,
whereby ${\cal P}_l$ stands for a set of numbers which give the
topology of the string. We will choose ${\cal P}=3D(\bbox{e}_1,
\bbox{e}_2, \dots)$, where the unit vector $\bbox{e}_i \in \{ \pm
\bbox{e}_x,\pm\bbox{e}_y\}$ gives the direction of the $i^{th}$
hop which the hole has taken. For the wave function of a hole
trapped in the vicinity of the site $l$ we then make the ansatz:
\begin{equation}
| \Psi_{l}\rangle  =3D \sum_{\{{\cal P}_l\}}
\alpha_{l,{\cal P}_l} |l,{\cal P}_l \rangle .
\label{pol}
\end{equation}
For the Hamiltonian we choose the sum of the kinetic energy and
Ising part of the Heisenberg exchange:
\begin{equation}
H_0 =3D -t \sum_{\langle i,j\rangle} \sum_\sigma
(\hat{c}_{i,\sigma}^\dagger\hat{c}_{j,\sigma} + H.c.) + J
\sum_{\langle i,j\rangle}\left( S_i^z S_j^z  -\frac{\hat{n}_i
\hat{n}_j}{4} \right).
\end{equation}
The kinetic energy has matrix elements of $-t$ between two states
$|l,{\cal P}_l \rangle$ and $|l,{\cal P}_l' \rangle$ if the
respective paths can be transformed into each other by a single
hop. The diagonal matrix elements of $H_0$ originate from the term
$J\sum_{\langle i,j \rangle} \left( S_{i,z} S_{j,z}-n_i n_j/4
\right)$. If we choose the state $|\Phi_0\rangle$ as the zero of
energy for these diagonal matrix elements, the energy of any given
string state is simply the number of additional `broken' bonds
times $J/2$. For example the energy of the state
$\hat{c}_{(0,1),\uparrow}|\Phi_0\rangle$ is $J/2$, while for the
state $\hat{c}_{(-1,1),\uparrow}|\Phi_0\rangle$ it is $3J/2$. \\
Once we truncate the subspace of string states (for example by
taking into account only paths up to a maximum length) we can
easily set up the matrix corresponding to $H_0$ and diagonalize it
numerically to obtain the ground state
\begin{equation}
H_0 | \Psi_l \rangle  =3D
E_l | \Psi_l \rangle .
\label{Sch0}
\end{equation}
This gives us the coefficients $\alpha_{l,{\cal P}_l}$ in
(\ref{pol}) as well as the ($l$-dependent) ground state energy
$E_l$. This procedure can be performed for any `environment' which
the hole has been created in. The variation of the wave function
$| \Psi_l \rangle $ with $l$ describes how the spin bag is
`deformed' by its local environment. This `prediagonalization' of
$H_0$ reduces the number of degrees of freedom considerably, and
thus makes the problem tractable even with the relatively large
unit-cell for the stripe structure. This is also the main
difference compared to the related method of
Trugman\cite{Trugman}, which obtains the ground state wave
function by constructing one Bloch states from each of the string
states $|l, {\cal P}\rangle$. Neglecting the excited states of
$H_0$ in the further steps of the calculation is justified
provided the separation in energy between these excited states and
the ground state is sufficiently large so that transitions into
these states can be neglected. We have verified that the
separation in energy between the ground state and the first
excited state is in any case $\ge J/2$, in most cases even larger
than $J$. Since the bandwidth for a single hole is $\approx 2J$,
neglecting these excited states is probably not really an
excellent but reasonable approximation.\\ In our calculation we
have used the part of the Hilbert space defined by the condition
$S_z=3D-1/2$, where $S_z$ is the z component of the total spin. The
other possible choice $S_z=3D1/2$ would give identical results,
corresponding to a hole moving on the opposite sublattice. The
condition $S_z=3D-1/2$ may be fulfilled by removing an up spin from
the stripe structure. There are 14 non-equivalent spin-up sites in
the elementary cell in Figure \ref{fig1}. The presence of the
ferromagnetic bonds across the stripes brings about an extra
complication: some of states that represent holes created in
stripes are directly coupled by the `fast' nearest neighbor
hopping term. For example, the state
$\hat{c}_{(5,1),\uparrow}|\Phi_0\rangle$ may be obtained by acting
only once with the operator $T_{\ldots}$ on
$\hat{c}_{(4,1),\uparrow}|\Phi_0\rangle$. This cannot happen in
the N\'eel state, where sites of like spin are always separated by
at least $2$ lattice spacings, and thus are not coupled by one
nearest neighbor hop. To deal with this situation, we incorporate
both states, $\hat{c}_{(5,1),\uparrow}|\Phi_0\rangle$ and
$\hat{c}_{(4,1),\uparrow}|\Phi_0\rangle$, into one and the same
self-trapped state $|\Psi_{l}\rangle$, which consequently is
elongated in $x$-direction. In this way the number of
non-equivalent starting points $l$ for holes is reduced to the
following 12 sites: $(0,1)$, $(1,0)$, $(1,2)$, $(2,1)$, $(2,3)$,
$(3,0)$, $(3,2)$, $(4,1)$, $(5,3)$, $(6,0)$, $(6,2)$ and $(7,3)$.
Finally, it turns out that some starting points $l$ for holes are
equivalent in that the respective spin polaron states $|
\Psi_{l}\rangle$ may be transformed into one other by applying a
symmetry transformation of the stripe structure. For example, the
sites $(0,1)$ and $(4,1)$ which belong to the same elementary cell
are equivalent.  \\ By means of computer algebra we can now create
the states $|\Psi_{l}\rangle$. The algorithm thereby takes into
account exactly strings of length up to 3 lattice spacings. For
longer paths (which are less relevant) the approximation is made
that their amplitude in $|\Psi_{l}\rangle$ does not depend on
details of further steps.  At this stage of the calculation we
also make the assumption that the states $| l, {\cal P}_l \rangle$
for different $l$ and/or different ${\cal P}_l$ are always
orthogonal. For longer paths this may not be true. The stripe
pattern provides much more possibilities to reach identical states
from different original configurations, not only by means of
retracing paths as it is basically in the case of the N\'eel
state, but we neglect this here. As already mentioned, any charge
and spin fluctuations of the holes and spins which form the
background stripe pattern are also neglected. \\ It then turns out
that the differences between the eigenenergies $E_l$ of
spin-polaron states for different initial positions $l$ of the
hole are $\sim J$.  The states which are lowest in energy thereby
correspond to polarons centered on $(0,1)$, $(4,1)$ and equivalent
positions. We see that despite the fact that the freedom of motion
of a hole created at the site $(0,1)$ is strongly reduced by the
(static) holes at $(0,0)$ and $(0,2)$, which inevitably results in
a loss of kinetic energy, there is a net gain in energy compared
to other sites. This is obviously due to the smaller number of
broken bonds created in the first few hops. This means first of
all a lower exchange energy, and second a gain in kinetic energy
because the hole can move more freely along the remaining paths.
We do not pursue this any further, but we note that this may be
one reason for the stability of stripes.
\section{Propagation of spin polarons in the stripe structure}
So far we have constructed wave function for self-trapped states
at various positions in the stripe structure. We now assume that
the wave function for a hole added to the stripe structure is a
coherent combination of these polaron states $|\Psi_{l}\rangle$.
All processes which were neglected when constructing the
self-trapped states $|\Psi_{l}\rangle$ will now be incorporated
into an effective Hamiltonian $H_{eff}$ which couples the
$|\Psi_{l}\rangle$ centered on different $l$. As already
mentioned, in doing so we implicitly assume that the excited
states which come out of (\ref{pol}) are separated from the ground
state by an energy which significantly exceeds the matrix elements
of $H_{eff}$.\\ The diagonal element $\langle \Psi_{l} | H_{eff}
|\Psi_{l}\rangle$ is essentially the eigenenergy $E_l$ of the
polaron-state $|\Psi_{l}\rangle$. It influences the probability
that a particular site is occupied by the polaron. By means of
computer algebra we have also set up additional diagonal and
off-diagonal contributions to $H_{eff}$. Thereby we have confined
the search to paths no longer than 2 lattice spacings. There are
several hundreds of process which couple pairs of sites one of
which (at least) belongs to the elementary cell. They form 29
different categories of processes. We shall now discuss some
examples. \\ To begin with, the possibility of direct hopping to
second nearest neighbors by virtue of the $t'$-term in the $t-J$
model has not been taken into account when setting up the
different $H_0$. Therefore, $H_{eff}$ must take it into account.
Consider Fig.\ref{fig2}. Shifting a hole created at the site
$(0,1)$ by $(1,1)$ gives rise to a state that represents a hole
created at $(1,2)$. This process therefore couples the states
$|\Psi_{(0,1)}\rangle$ and $|\Psi_{(1,2)}\rangle$, the numerical
value of the corresponding matrix element $\langle \Psi_{(0,1)} |
H_{eff} |\Psi_{(1,2)}\rangle$ is  $t^\prime
\alpha_{(0,1)}\alpha_{(1,2)}$. Here pairs of numbers in subscripts
refer to position of holes, and the $\alpha$'s are obtained by
diagonalizing the $H_0$ matrices.\\ Next, Fig.\ref{fig2}(a,b,c)
depicts a process which contributes to the diagonal element for
the spin-polaron state at $(2,1)$. After a single hop in the
$(1,0)$ direction a hole created at $(2,1)$ occupies the site
$(3,1)$. The amplitude of this state in $|\Psi_{(2,1)}\rangle$ is
$\alpha_{l,{\cal P}_l}=3D\alpha_{(2,1),(1,0)}$, where, as explained
above, the first pair of numbers denotes the original position of
the hole whereas the second pair gives the direction of the hop.
Next, let us assume that the $t^\prime$ term now shifts the hole
to the site $(2,2)$. This state has the coefficient
$\alpha_{(2,1),(0,1)}$ in $|\Psi_{(2,1)}\rangle$ whence we have
found a contribution  of $t^\prime \alpha_{(2,1),(1,0)}
\alpha_{(2,1),(0,1)}$ to the matrix element $\langle \Psi_{(2,1)}
| H_{eff} |\Psi_{(2,1)}\rangle$.\\ There are a few more types of
processes that involve the $t^\prime$ term. In Fig.\ref{fig2}(d)
we start with the string state obtained by creating a hole at
$(2,3)$ and performing two hops in directions $(1,0)$ and $(0,1)$.
Shifting the hole in $(-1,-1)$ direction by virtue of $t'$, this
is transformed into a state obtained by creating a hole at $(3,4)$
and shifting it in the directions $(0,-1)$ and $(-1,0)$
(Fig.\ref{fig2}(e)). This type of process couples polarons at the
sites $(2,3)$ and  $(3,4)$ and the corresponding contribution to
$\langle \Psi_{(2,3)} | H_{eff} |\Psi_{(3,4)}\rangle$ is $t^\prime
\alpha_{(2,3),(1,0)(0,1)} \alpha_{(3,0),(0,-1)(-1,0)}$.
Yet another process is possible due to the
existence of parallel spins on nearest neighbor sites in the stripe
structure. Shifting  a hole created at the site  $(3,2)$
(see Fig.\ref{fig2}(f)) by $(1,1)$ produces the same state as obtained =
by
two hops of this hole in the directions $(0,1)$ and
$(1,0)$. The corresponding
contribution to the matrix element
$\langle \Psi_{(3,2)} | H_{eff} |\Psi_{(3,2)}\rangle$ is $t^\prime
\alpha_{(3,2)} \alpha_{(3,2),(0,1)(1,0)}$. \\
The third nearest neighbor
hopping term $t''$ model gives rise to similar additions to
$H_{eff}$. For example a contribution $t^{\prime \prime} \alpha_{(5,3)}
\alpha_{(7,3)}$  to   $\langle \Psi_{(5,3)} | H_{eff}
|\Psi_{(7,3)}\rangle$ emerges due to the possibility of direct hopping =
of
a hole from $(5,3) \rightarrow (7,3)$. Fig.\ref{fig3}
shows a process whereby a hole created at the site
$(4,5)$ (Fig.\ref{fig3}(a))  which hopped three times in the
directions $(1,0)$, $(0,-1)$ and $(0,-1)$ (Fig.\ref{fig3}(b))
is shifted by $(0,2)$.
The resulting state (Fig.\ref{fig3}(c)) then represents
a hole created at $(5,3)$ which hopped twice in direction $(0,1)$.
That gives rise to a contribution $- t^{\prime \prime}
\alpha_{(4,5),(1,0)(0,-1)(0,-1)} \alpha_{(5,3),(0,1)(0,1)}$ to the
matrix element $\langle \Psi_{(4,5)} | H_{eff}
|\Psi_{(5,3)}\rangle$. The minus sign stems from Fermi statistics
and the definition of the operator $T_{\langle i,j \rangle}$ which
creates string states and the fact that the length of the first
string is odd, while the length of the second is even. \\ This
concludes our discussion of the effects of $t'$ and $t''$ and we
now proceed to a discussion of contributions to $H_{eff}$ which
originate from the exchange part of the $t-J$ model.
Fig.\ref{fig4}(a,b,c,d) depicts a process that involves pairs of
parallel spins on nearest neighbor sites in the stripe structure.
Fig.\ref{fig4}(a,b) shows holes which have been created at the
sites $(4,1)$ and $(8,1)$. After three hops in the direction
$(-1,0)$ the hole from the site $(8,1)$ reaches the site $(5,1)$.
Fig.\ref{fig4}(c) shows the corresponding string state which is a
component of the spin polaron at the site $(8,1)$. By acting with
the spin-flip term on the bond $(6,1)-(7,1)$ this state is
transformed into the state in Fig.\ref{fig4}(d), which in turn is
a string state which is a component of the spin polaron at the
site $(4,1)$. To be more precise, the state in Fig.\ref{fig4}(d)
can be obtained by a single hop in the $(1,0)$ direction from the
state in Fig.\ref{fig4}a. The matrix element describing the
process in Fig.\ref{fig4}(a,b,c,d) is $(J/2) \alpha_{(4,1),(1,0)}
\alpha_{(8,1),(-1,0)(-1,0)(-1,0)}$. This mechanism of coupling
spin polarons at the sites $(4,1)$ and $(8,1)$ also works  for all
longer string states where the hole in Fig.\ref{fig4} (c) and (d)
has moved in the vertical direction. This motion does not
interfere with the action of the exchange term on the sites
$(6,1)$ and $(7,1)$ which gives rise to the equivalence of string
states formed by further motion of the hole in Fig.\ref{fig4}(c)
and (d). The above contribution to $H_{eff}$ therefore should be
supplemented by terms related to all relevant longer paths. \\
Quite generally it is obvious that Figures \ref{fig2},\ref{fig3},
and \ref{fig4} represent only a small fraction of all processes
which may occur during the motion of a hole through the stripe
background. There are various inequivalent positions of polarons
within the unit cell and the different local arrangement of spins
and holes in the stripes gives rise to many non-equivalent
processes which however may be easily found by means of computer
algebra.

As we have mentioned before, spin and charge fluctuations have been
neglected in our calculation. They influence to some extent the =
properties
of the system. Nevertheless, it seems plausible as a first attempt to
treat the stripe structure as a static object and discuss the =
propagation
of an excess hole in the stripe background. We expect to gain in this
way some insight into electronic properties of stripes.

In order to construct a coherent combination of spin polaron states
we have to label the positions of polarons in a slightly different way. =
We
identify a site $l$ by the vector ${\bf R}_m$ which gives the
position of the lower left corner of the elementary cell containing
the site and the position $n$ of the site within the elementary cell:
\begin{equation}
|\Psi_{l}\rangle \equiv |\Psi_{{\bf R}_m,n}\rangle
\end{equation}
A Bloch state $|\tilde{\Psi} \rangle$ with momentum ${\bf P}$ then
can be written as
\begin{equation}
|\tilde{\Psi}_n \rangle =3D\frac{1}{\sqrt{N^\prime}}
e^{i {\bf R}_m {\bf P}} \sum_m
|\Psi_{{\bf R}_m,n}\rangle,
\end{equation}
where $N^\prime$ is a normalization constant, i.e.
the number of elementary cells
in a system with periodic boundary conditions.
The momentum ${\bf P}$ has to be chosen from the
reduced Brillouin zone $[-\frac{\pi}{8},\frac{\pi}{8}]
\otimes [-\frac{\pi}{4},\frac{\pi}{4}]$.
The states $|\tilde{\Psi}_n \rangle$ are normalized  by definition:
\begin{equation}
\langle\tilde{\Psi}_{n} |\tilde{\Psi}_{n^\prime}
\rangle=3D\delta_{n,n^\prime}
\end{equation}
The states $|\tilde{\Psi}_{n^\prime} \rangle$ and $|\tilde{\Psi}_{n}
\rangle$ for different sites $n$ and  $n^\prime$ in the elementary cell
are orthogonal because up to the accuracy level of our calculation
different string states are orthogonal. The matrix elements of the
Hamiltonian for the coherent combinations $|\tilde{\Psi}_{n}\rangle$ =
then
may be
represented in terms of the matrix elements for spin polarons which we
already know:
\begin{equation}
\langle\tilde{\Psi}_{n}|H_{eff}|\tilde{\Psi}_{n^\prime} \rangle=3D
\sum_{m^\prime}  e^{i {\bf R}_{m^\prime} {\bf P}}
\langle \Psi_{0,n} |H_{eff}| \Psi_{{\bf R}_{m^\prime},n^\prime}\rangle.
\label{ham}
\end{equation}
In order to get a feeling for the evolution with doping we have
analyzed an analogous stripe structure, shown in Figure
\ref{fig5}, which would correspond to the case of 1 hole per 12
copper sites. The existence of stripes at this level of doping is
as yet a speculation. Stripes, if they exist at all, may in this
case have a dynamical character. Nevertheless, we have decided to
use a structure which is analogous to the pattern that we used for
the $1/8$ case. This can also give some indication as to how
robust the various features in the calculated spectral density are
against possible changes of the structure. Finally, to conclude
the discussion of the propagation of the mobile hole, we note that
the dispersion relation for a hole in the original N\'eel state,
can be carried out in precisely the same way\cite{ederbecker}. We
have also performed this calculation to compare the dispersion for
the N\'eel state and the stripe phase.
\section{Spectral function}
We now proceed to discuss the one-particle spectral function and
consider the term related to creation of an additional hole in the
stripe structure. It can be written as
\begin{equation}
A_h({\bf k},\omega) =3D \sum_{\nu}
|\langle \Phi_{\nu}^{(+1h)} | c_{{\bf k},\uparrow}
| \Phi_{0}\rangle |^2 \delta\left(\omega -
(E_{\nu}^{(+1h)}-E_{0}^{(s)})\right),
\end{equation}
where $|\Phi_{\nu}^{(+1h)}\rangle$ is an eigenstate corresponding to an
additional hole inserted into the stripe structure and in our
calculation is identified  with the solution of the eigenvector
problem for the Hamiltonian (\ref{ham}).
The $c_{{\bf k},\uparrow}$ operator  may be written as
\begin{equation}
c_{{\bf k},\uparrow}=3D\frac{1}{\sqrt{N}} \sum_n e^{-i {\bf k} {\bf R}_n =
}\
c_{n,\uparrow}.
\end{equation}
The summation here runs over all sites in the lattice and $N=3D32N'$ is =
their
total number.
The $\nu$-th excited state $|\Phi_{\nu}^{(+1h)}\rangle$ is a linear
combination of the states $|\tilde{\Psi}_{n}\rangle$:
\begin{equation}
|\Phi_{\nu}^{(+1h)}\rangle =3D\sum_{n}  \beta^{(n)}_{\nu}
|\tilde{\Psi}_{n}\rangle.
\end{equation}
The summation label $n$ thereby
refers to the 12 inequivalent sites in the elementary cell enumerated
above, which
may be occupied by a spin-up polaron. After some straightforward algebra =
the
relevant correlation function takes the simple form
\begin{eqnarray}
&&\langle \Phi_{\nu}^{(+1h)}|c_{{\bf k},\uparrow}
| \Phi_{0}^{(s)}\rangle =3D \delta_{Mod(P_x+k_x,2\pi/8),0}
\times \nonumber \\
&& \delta_{Mod(P_y+k_y,2\pi/4),0}
\sqrt{\frac{N^\prime}{N}} \sum_{n,l}
(\beta^{(n)}_\nu)^{\ast} e^{-{\bf k} {\bf R}_l } \langle
\Psi_{0,n}|c_{l,\uparrow}
| \Phi_{0}\rangle.
\label{corfun}
\end{eqnarray}
The matrix elements $\langle \Psi_{0,n}|c_{l,\uparrow} | =
\Phi_{0}\rangle$
give the overlap between the bare hole created in the stripe structure =
at
site $l$ and
the self-trapped state centered on site $n$. Since the electron
annihilation operator cannot create any spin defects, it can only
produce the `string of length zero'. Hence only
the state $c_{n,\uparrow} | \Phi_{0}\rangle$, which
has the prefactor $\alpha_{n}$ in
$|\Psi_{0,n} \rangle$, can possibly have a nonvanishing overlap with
$c_{l,\uparrow} | \Phi_{0}\rangle$. Therefore we have
$\langle \Psi_{0,n}|c_{l,\uparrow} | \Phi_{0}\rangle=3D\delta_{l,n}
\alpha_n$. The only exception are the `elongated' states
along the ferromagnetic bonds. As was explained above,
the states obtained by creating a hole either at $(4,1)$ or at $(5,1)$
are both included into the self-trapped state centered on $(4,1)$.
Hence
\begin{equation}
\langle \Psi_{(4,1),0}|c_{l,\uparrow} | \Phi_{0}\rangle
=3D \delta_{l,(4,1)} \alpha_{(4,1)}
- \delta_{l,(5,1)} \alpha_{(4,1),(1,0)},
\end{equation}
where the relative minus between the two contributions is due to
Fermi statistics.\\ We have employed the formula (\ref{corfun}),
together with the above expressions for the matrix elements, in
the evaluation of the spectral functions. It should be noted that
the resulting `spectral weight' basically is a measure as to how
well the respective many-body wave function matches the wave
function of a photoelectron, which we model by a plane wave with
momentum ${\bf k}$. The actual spectral weight observed in an
angle resolved photoemission experiment may still differ from
this: the spectral weight of a quasiparticle in the t-J model is
known to have a significant ${\bf k}$ dependence due to the
quantum spin fluctuations of the Heisenberg
antiferromagnet\cite{spec}, in the more realistic
single\cite{henk} or three-band\cite{aligia1} Hubbard model an
additional ${\bf k}$ dependence is brought about by charge
fluctuations and/or interference between photoholes created on Cu
and O sites\cite{aligia1}. Unfortunately these effects are beyond
the scope of our approximation.\\ On the other hand, the Brillouin
zone of the supercell in Figure \ref{fig1} is
$[-\frac{\pi}{8},\frac{\pi}{8}] \otimes
[-\frac{\pi}{4},\frac{\pi}{4}]$. An angle resolved photoemission
experiment with a momentum transfer of, for example,
$\bbox{k}=3D(\pi,0)$ thus actually probes the eigenvalue spectrum at
$(\frac{\pi}{8},0)$. Any `dispersion' which is measured on scales
in $\bbox{k}$ space larger than the extent of the reduced
Brillouin zone therefore is in reality a `spectral weight
dispersion' and cannot be interpreted as a band dispersion. This
`dispersion' then is essentially determined by the expression
above, namely how well a photoelectron with the respective
momentum matches the wave function of the respective state within
the 32-site unit cell. In that sense the above `spectral weight'
does very well have some significance. Therefore, one may expect
that although the magnitude of the spectral weight obtained from
(\ref{corfun}) cannot be directly compared to experiments, it
still may give us a rough feeling as to where in $({\bf
k},\omega)$ space we may expect nonvanishing weight.\\ The
smallness of the reduced Brillouin zone brings about another
complication: suppose we compute (as is usually done) the spectral
function at an equally spaced $\bbox{k}$-grid along some
high-symmetry line, e.g. $(0,0)\rightarrow (\pi,0)$. Then, there
are two possibilities: the step size of the $\bbox{k}$-grid may be
commensurate with the size of the reduced Brillouin zone or not.
To illustrate the first case, let us assume the step size to be
$\pi/8$. Then, we would actually be probing just two momenta in
the reduced zone, namely $(\pi/8,0)$ and $(0,0)$. This would
therefore give us an artificial periodicity of the peak positions
(although not of the spectral weight). On the other hand, if we
choose an incommensurate step size (we choose $\pi/7$ as an
example) we would walk through the reduced Brillouin zone like
this: $(0,0)$, $(-\frac{6\pi}{56},0)$, $(\frac{2\pi}{56},0)$,
$(-\frac{4\pi}{56},0)$, $(\frac{4\pi}{56},0)\dots$ that means more
or less in a random sequence. Plotting the results for the
spectral density in this way therefore is probably of very little
significance. We have therefore decided to average the spectra for
each momentum over a small neighborhood of the respective momentum
and compute:
\begin{equation}
\bar{A}(\bbox{k},\omega) =3D \sum_{i,j}
w_{i,j} A_h(\bbox{k}+ i\delta_x + j\delta_y,\omega).
\end{equation}
We have chosen the weight function $w_{i,j}$ to be a constant and
used $\delta_{x,y}=3D\frac{\pi}{8\cdot 16} \bbox{e}_{x,y}$, and
summed over $-8 \leq i,j \leq 8$. To some degree this also
simulates the effect of the finite $\bbox{k}$-resolution in an
ARPES spectrum. Bearing this in mind we turn to the numerical
results for the spectral density. Figures \ref{fig6}-\ref{fig10}
then show the spectral function $\bar{A}(\bbox{k},\omega)$
calculated along various high-symmetry lines in the {\em extended}
Brillouin zone. Distances in the $k$ space between points in the
sequence are equal.  An important point for comparison with
experiment is the following: we are using a hole language, which
implies that the ground state of the hole is the state with the
most negative energy. In a photoemission experiment the
single-hole ground state then would actually be the first
ionization state, i.e. the state with the lowest binding energy,
and would form the `top of the band'. For comparison with the
usual way of plotting experimental ARPES data, with binding energy
increasing to the left, the energy axis in our figures would have
to be inverted. To begin with, the direction $(0,0)\rightarrow
(\pi,\pi))$ (see Figure \ref{fig10}) and $(0,0)\rightarrow
(0,\pi)$ (which has momentum parallel to the stripes, see Figure
\ref{fig7}) show a distribution of spectral weight which broadly
follows the quasiparticle dispersion for the $t-t'-t''-J$ model in
the pure antiferromagnet (the latter is indicated by the full
line). The direction $(\pi,0)\rightarrow (\pi,\pi)$ (see Figure
\ref{fig8}) might be interpreted as the superposition of two
components, one of them following the antiferromagnetic
quasiparticle dispersion, the other one forming a dispersionless
band at $\approx -2.3 t$. It should be noted, however, that in
this part of the extended Brillouin zone the spectral weight of
the quasiparticle band is practically zero in the
antiferromagnet\cite{Wellsetal}, whence the dispersionless band is
probably unobservable. The situation is different for the
direction $(0,0)\rightarrow (\pi,0)$ (i.e. the momentum is
perpendicular to the stripes), see Figure \ref{fig6}. In this
direction we see several more or less dispersionless `bands', and
in particular a flat band of low energy states at $\approx -2.3
t$. Since this range of momenta is within the antiferromagnetic
Brillouin zone one may expect, based on the results for the
antiferromagnet, that it has an appreciable spectral weight. A
dispersionless band at $\approx -2.3t$ is also seen in the
direction $(0,\pi)\rightarrow (\pi,\pi)$, but again thus is
probably not observable because it is in the outer part of the
extended Brillouin zone. 
Averaging the directions $(0,\pi)$ and $(\pi,0)$, which presumably
corresponds to the experimental situation because the stripes do
not have a uniform direction throughout the sample, one would thus
get a `broadened' version of the quasiparticle band in the
insulator, plus a relatively flat band of new states, which
actually form the first ionization states. It is quite tempting to
identify this with the experimental results of Ino {\em et al.}.
At half-filling $(x=3D0)$, these data showed a dispersion which is
very much reminiscent of the `quasiparticle band' observed in
Sr$_2$CuO$_2$Cl$_2$. As doping is increased, this band persists to
some degree, although is getting more diffuse. In addition, around
$(\pi,0)$ a relatively dispersionless band of `new states'
emerges, which forms the first ionization states. 
In order to clarify the character of the wave functions, we have
also studied the real space wave function
$w_n=3D|\beta^{(n)}_{\nu}|^2$ for different states (see Eq. (12)).
This basically gives the probability to find the hole in the
self-trapped state at site $n$ within the unit cell.
Figure \ref{fig11} shows this for a state which produces an
intense peak at $(\pi,0)$, whereby the peak relatively far from
the top of the band and thus would correspond to the `remnant of
the antiferromagnetic band'. It is obvious from the figure, that
the mobile hole in this case avoids the hole stripes and resides
mainly in the center of one of the antiferromagnetic domains. On
the other hand, Figure \ref{fig12} shows the probability
distribution for one of the `new states' at $(\pi,0)$. There the
extra hole resides almost exclusively within the stripe. This
shows again, that the new states, which appear upon doping, are
related to states which are predominantly localized in the hole
stripes. Of course one has to keep in mind that our approximation
of static holes within the stripes will make the description of
precisely these states the most questionable. Finally, Figure
\ref{fig13} shows $w_n$ for a state which produces a significant
peak at $(0,\pi)$ which also belongs to the `remnant of the
antiferromagnetic band'. Again, we see that the mobile hole avoids
the hole stripes, which is expected if the state somehow resembles
properties of the undoped antiferromagnet. 
In order to understand the origin of some features observed in the
spectra for the $1/8$ case we have also performed the calculation
of the spectral function for the structure represented by Fig.
\ref{fig5}, which would correspond to the case of $\delta =3D 1/12$.
Stripes at this level of doping, if exist, have a dynamical
character and the applicability of the stiff structure represented
by Fig. \ref{fig5} is very questionable. Our aim was to check the
robustness of some characteristics, rather than to derive any
spectra comparable with experiments. Figures \ref{fig14} and
\ref{fig15} depict spectra calculated along the lines $(0,0)
\rightarrow (\pi,0)$ and $(0,0) \rightarrow (\pi,\pi)$
respectively. We again notice a band at the energy $\approx -2.3t$
which  emerges in the vicinity of $(\pi,0)$ and disappears
somewhere halfway to the point $(0,0)$. We have checked that this
band may be attributed to states that represent the extra hole
which resides predominantly within the stripe. The rest of the
spectrum seems to  resemble the energy dispersion of a single hole
in  an antiferromagnet. We have not observed that for the $1/8$
case when the size of antiferromagnetic domains is smaller. The
states which form the `antiferromagnetic band' correspond to the
extra hole avoiding stripes.  The spectrum for the direction
$(0,0) \rightarrow (\pi,\pi)$, however very broad, follows the
quasiparticle dispersion of the pure antiferromagnet in a more
evident way then for  $\delta=3D1/8$. These remarks indicate that
the evolution of the spectral function between doping levels
$\delta=3D0$ and $\delta=3D1/8$ may be attributed to formation of
stripes.

\section{Discussion}
In summary, we have calculated the dispersion for a single mobile
hole in a static stripe structure or, put another way, a depleted
Heisenberg plane, thereby using a reasonably realistic version of
the $t-t'-t''-J$ model. The strongest approximation presumably
consisted in taking the holes in the stripes to be static, but we
believe this is a reasonable first step in understanding the
dynamics of holes in truly fluctuating stripes. In particular, the
present calculation incorporates the effect of the oscillating
direction of N\'eel order in the undoped antiferromagnetic domains
of the stripe pattern, as well as the influence of the important
$t'$ and $t''$ terms on the hole motion. From the calculated
dispersion and wave function we have computed the `spectral
weight' as a function of energy. This spectral weight is basically
the overlap of the wave function in the supercell with a plane
wave of the respective momentum. In an ARPES experiment on a
system with a relatively small reduced Brillouin zone as the
stripe structure with its large unit-cell, it is basically this
`spectral weight dispersion' which is being measured.\\ As
expected, the primary influence of the stripes is the `broadening'
of the spectral weight, which however still roughly follows the
hole dispersion in the insulator, at least along some
high-symmetry lines. This is to be expected for a not too strong
super-cell potential. In addition, the stripes lead to the
emerging of `new states' near $(\pi,0)$ at relatively low binding
energy. More precisely, the spectral density in the stripe
structure shows a rather sharp and dispersionless low energy band,
which is presumably most intense  around $(\pi,0)$, and which
actually forms the first ionization states. Analysis of the wave
functions has shown, that the holes reside predominantly within
the hole stripes in these new states, whereas the states
responsible for the `remnants of the antiferromagnetic band' in
the spectral density have the holes predominantly in the
antiferromagnetic domains. It is interesting to note that this
effect, namely the generation of low energy states at $(\pi,0)$ by
going from the undoped to the doped system, can also be observed
in exact diagonalization calculations for the $t-t'-t''-J$
model\cite{EdOh3}. Whether this is due to formation of stripes
even in the relatively small clusters studied remains to be
clarified.\\ In any case the effect of `doping' (that means the
modification of the N\'eel state to the stripe pattern with static
holes and antiferromagnetic domains) on the spectral function in
our calculation shows some similarity with recent ARPES
measurements on La$_{2-x}$ Sr$_x$CuO$_4$ \cite{Inoetal}. They
covered the range from an optimally doped superconductor
($x=3D0.15$) to an antiferromagnetic insulator (x=3D0). It has been
suggested on the basis of the incommensurate peaks in inelastic
neutron scattering \cite{Tranquadaetal,Bianconietal} that the
family of LSCO systems reveals an instability towards a {\em
static} spin-charge order of stripe form. The formation of these
stripe structures is accompanied by a suppression of $T_c$ at a
hole concentration $\delta \simeq 1/8$. \\ It should be noted that
in the experimental ARPES spectra `peaks' in the spectral density
cannot usually be assigned. The various `bands' in the spectra
become visible only upon forming the 2$^{nd}$ derivative of the
spectral density with respect to energy. For the insulating
compound La$_2$CuO$_4$ the `dispersion' roughly resembles the
known band structure of a hole in an antiferromagnet as it has
been observed in Sr$_2$CuO$_2$Cl$_2$: the top of the band appears
to be at $(\pi/2,\pi/2)$, the band at $(\pi,0)$ is at a
significantly ($\approx 200meV$) higher binding energy. Upon
doping this band structure of the insulator persists to some
degree, but in addition `new states' at $(\pi,0)$ appear. These
have a very low binding energy and indeed do form the first
ionization states. Let us concentrate on the case of $x=3D0.12$
which is closest to the configuration we have discussed. It is
believed that stripe structures are perpendicular in nearest
copper-oxygen planes. The spectra in the $(0,0) \rightarrow
(0,\pi)$ direction are a composition of spectra parallel and
perpendicular to stripes. The measurements in that direction
correspond rather well to a combination of our calculated spectra
for the directions $(0,0) \rightarrow (0,\pi)$ and $(0,0)
\rightarrow (\pi,0)$. The same holds true for the direction
$(\pi,0) \rightarrow (\pi,\pi)$. Keeping that in mind we notice
that the experimental and theoretical spectra show at least a
qualitative agreement.  In both cases we observe, along the
directions $(0,0) \rightarrow (\pi,0)$ and $(0,0) \rightarrow
(0,\pi)$, the `splitting' of the spectrum and formation of new low
energy states around $(\pi,0)$. This splitting is much less
pronounced in the direction $(0,\pi) \rightarrow (\pi,\pi)$. The
spectra along the direction $(\pi,0) \rightarrow (\pi,\pi)$ are
very broad. That prevents assignment of a band for some range of
${\bf k}$ points in that direction. Finally, the spectra along the
direction $(0,0) \rightarrow (\pi,\pi)$ are very broad and it is
practically impossible to distinguish any band.  \vfil Useful
discussion with J. Zaanen are gratefully acknowledged. One of the
authors (P.W.) acknowledges support by the Polish Science
Committee (KBN) under contract No. 2PO3B-02415.

\begin{figure}
\caption{Stripe structure for $12.5\%$ of doping which have been applied =
in the
calculation of the hole dispersion. Solid circles depict immobile
holes within the stripes, arrows
represent spins in antiphase AF domains that separate the stripes.}
\label{fig1}
\end{figure}
\begin{figure}
\caption{Some processes related to direct hopping  to next nearest
neighbors.}
\label{fig2}
\end{figure}
\begin{figure}
\caption{A process related to direct hopping  to third nearest
neighbors.}
\label{fig3}
\end{figure}
\begin{figure}
\caption{A processes related to the exchange term.}
\label{fig4}
\end{figure}
\begin{figure}
\caption{Structure applied in the calculation for the case of 1 hole
per 12 sites.}
\label{fig5}
\end{figure}
\begin{figure}
\caption{Spectral functions in the  $(0,0) \rightarrow (\pi,0)$
direction for  points separated by an equal distance.}
\label{fig6}
\end{figure}
\begin{figure}
\caption{Spectral functions in the  $(0,0) \rightarrow (0,\pi)$
direction for  points separated by an equal distance.}
\label{fig7}
\end{figure} 
\begin{figure}
\caption{Spectral functions in the  $ (\pi,0) \rightarrow (\pi,\pi)$
direction for  points separated by an equal distance.}
\label{fig8}
\end{figure}
\begin{figure}
\caption{Spectral functions in the
direction $(0,\pi)  \rightarrow (\pi,\pi)$  for points separated by an =
equal distance.}
\label{fig9}
\end{figure}
\begin{figure}
\caption{Spectral functions in the
direction $(0,0)  \rightarrow (\pi,\pi)$  for points separated by an =
equal distance.}
\label{fig10}
\end{figure} 
\begin{figure}
\caption{Quantum mechanical probability $w$ that an additional
hole is in the polaron state at  a particular site in the stripe
structure for a wave function related to a dominant peak in the
spectral function at the energy about $-1.60 t$ for ${\bf
p}=3D(\pi,0)$.} \label{fig11}
\end{figure}
\begin{figure}
\caption{Quantum mechanical probability $w$ that an additional
hole is in the polaron state at the site $n$ in the stripe
structure for a wave function related to a recognizable peak in
the spectral function at the energy about $-2.23 t$ for ${\bf
p}=3D(\pi,0)$. The hole stripes are in the 0$^{th}$ and 4$^{th}$
column.} \label{fig12}
\end{figure}
\begin{figure}
\caption{Quantum mechanical probability $w$ that an additional
hole is in the polaron state at a  particular site in the stripe
structure for a wave function related to a dominant peak in the
spectral function at the energy about $-1.63 t$ for  ${\bf
p}=3D(0,\pi)$.} \label{fig13}
\end{figure}
\begin{figure}
\vspace{0.3cm} \caption{Spectral functions in the  $(0,0)
\rightarrow (\pi,0)$ direction for  points separated by an equal
distance. Calculation was performed for the structure depicted in
Fig. \ref{fig5}.} \label{fig14}
\end{figure}
\begin{figure}
\vspace{0.3cm} \caption{Spectral functions in the direction $(0,0)
\rightarrow (\pi,\pi)$  for points separated by an equal distance.
Calculation was performed for the structure depicted in Fig.
\ref{fig5}. } \label{fig15}
\end{figure}

\end{document}